\documentstyle[12pt]{article}
\catcode`@=11
\newcount\@tempcntc
\def\@citex[#1]#2{\if@filesw\immediate\write\@auxout{\string\citation{#2}}\fi
  \@tempcnta\z@\@tempcntb\m@ne\def\@citea{}\@cite{\@for\@citeb:=#2\do
    {\@ifundefined
       {b@\@citeb}{\@citeo\@tempcntb\m@ne\@citea\def\@citea{,}{\bf ?}\@warning
       {Citation `\@citeb' on page \thepage \space undefined}}%
    {\setbox\z@\hbox{\global\@tempcntc0\csname b@\@citeb\endcsname\relax}%
     \ifnum\@tempcntc=\z@ \@citeo\@tempcntb\m@ne
       \@citea\def\@citea{,}\hbox{\csname b@\@citeb\endcsname}%
     \else
      \advance\@tempcntb\@ne
      \ifnum\@tempcntb=\@tempcntc
      \else\advance\@tempcntb\m@ne\@citeo
      \@tempcnta\@tempcntc\@tempcntb\@tempcntc\fi\fi}}\@citeo}{#1}}
\def\@citeo{\ifnum\@tempcnta>\@tempcntb\else\@citea\def\@citea{,}%
  \ifnum\@tempcnta=\@tempcntb\the\@tempcnta\else
   {\advance\@tempcnta\@ne\ifnum\@tempcnta=\@tempcntb \else \def\@citea{--}\fi
    \advance\@tempcnta\m@ne\the\@tempcnta\@citea\the\@tempcntb}\fi\fi}
\catcode`@=12

\def\theequation{\arabic{section}.\arabic{equation}}
\def\barr{\begin{array}}
\def\earr{\end{array}}
\def\beq{\begin{equation}}
\def\eeq{\end{equation}}
\def\bea{\begin{eqnarray}}
\def\eea{\end{eqnarray}}
\def\bmath{\begin{displaymath}}
\def\emath{\end{displaymath}}
\def\bq{\begin{quote}}
\def\eq{\end{quote}}
\def\Re{\mbox{Re}}
\def\Im{\mbox{Im}}

\def\cF{{\cal F}}
\def\cL{{\cal L}}
\def\cM{{\cal M}}
\def\cO{{\cal O}}
\def\cT{{\cal T}}
\def\li{\lambda_i}
\def\lj{\lambda_j}
\def\PL{\mbox{P}_L}
\def\PR{\mbox{P}_R}
\def\apprle{\hspace{-0.1cm}\stackrel{\displaystyle <}{\sim}}
\def\apprge{\hspace{-0.1cm}\stackrel{\displaystyle >}{\sim}}
\def\slash#1{\setbox0=\hbox{$#1$}#1\hskip-\wd0\hbox to\wd0{\hss\sl/\/\hss}}

\def\snG{\mbox{{\footnotesize n}}_G}
\def\snR{\mbox{{\footnotesize n}}_R}
\def\nG{\mbox{n}_G}
\def\nR{\mbox{n}_R}
\def\qmd{q_\mu}
\def\qnu{q^\nu}
\def\smnd{\sigma_{\mu\nu}}
\def\gmd{\gamma_\mu}
\def\gmu{\gamma^\mu}
\def\veps{\varepsilon}
\def\g5{\gamma_5}
\def\qsl{\slash{q}}
\def\lNi{\lambda_{N_i}}
\def\lNj{\lambda_{N_j}}
\def\lZ{\lambda_Z}
\def\Li2{\mbox{Li$_2$}}

\def\rpl{\rho_+}
\def\rmi{\rho_-}
\def\xpl{\xi_+}
\def\xmi{\xi_-}

\def\Abs{\mbox{Abs}}
\def\lN1{\lambda_{N_1}}

\def\arnpp#1{{\em Ann.\ Rev.\ Nucl.\ Part.\ Phys.\ }{\bf #1}}

\def\mpla#1{{\em Mod.\ Phys.\ Lett.\ }{\bf A#1}}

\def\npb#1{{\em Nucl.\ Phys.\ }{\bf B#1}}
\def\plb#1{{\em Phys.\ Lett.\ }{\bf A#1}}
\def\plb#1{{\em Phys.\ Lett.\ }{\bf B#1}}
\def\prl#1{{\em Phys.\ Rev.\ Lett.\ }{\bf #1}}

\def\prd#1{{\em Phys.\ Rev.\ }{\bf D#1}}
\def\prepc#1{{\em Phys.\ Rep.\ }{\bf C#1}}
\def\prep#1{{\em Phys.\ Rep.\ }{\bf #1}}
\def\ptp#1{{\em Prog.\ Theor.\ Phys.\ }{\bf #1}}

\def\zpc#1{{\em Z.\ Phys.\ }{\bf C#1}}
\begin{document}

\begin{flushright}
RAL/94-032\\[-0.2cm]
MZ-TH/94-15
\end{flushright}

\begin{center}
{\Large{\bf Flavour-Violating Charged Lepton Decays}}\\[0.4cm]
{\Large{\bf in Seesaw-Type Models }}\\[1.5cm]
{\large A. Ilakovac}$^{a}$\footnote[1]{On 
leave from Department of Physics, University of 
Zagreb, 41001 Zagreb, Croatia.}{\large ~and A. Pilaftsis}$^{b}$\footnote[2]{
E-mail address: pilaftsis@v2.rl.ac.uk}\\[0.4cm]
$^{a}$ {\em Institut f\"ur Physik, $THEP$,
Johannes-Gutenberg Universit\"at, 55099 Mainz, Germany}\\[0.3cm]
$^{b}$ {\em Rutherford Appleton Laboratory, Chilton, Didcot, Oxon, OX11 0QX, 
England}
\end{center}
\vskip1cm
\centerline {\bf ABSTRACT}
Analytic expressions of lepton-flavour- and lepton-number-violating
decays of charged leptons are derived in the context of general
$SU(2)_L\otimes U(1)_Y$ seesaw scenarios that are motivated by grand
unified theories (GUT's) or superstring models, in which left-handed
and/or right-handed neutral singlets are present. Possible constraints
imposed by cosmology and low-energy data are briefly discussed. The
violation of the decoupling theorem in flavour-dependent graphs due to
the presence of heavy neutral leptons of Dirac or Majorana nature is
emphasized. Numerical estimates reveal that the decays $\tau^-\to
e^-e^-e^+$ or $\tau^-\to e^-\mu^-\mu^+$ can be as large as $\sim
10^{-6}$, which may be observed in LEP experiments or other $\tau$
factories.

\newpage

\section{Introduction}
\indent

The quest for an understanding of the problem of smallness in mass or
masslessness of the light known neutrinos, $\nu_e$, $\nu_\mu$, and 
$\nu_\tau$, has relied on interesting solutions in the context of extended 
gauge structures of the minimal Standard Model (SM), such as grand unified 
theories, {\em e.g.} $SO(10)$ models~\cite{GUT}, or superstring 
models with an $E_6$ symmetry~\cite{EW}. Among the various solutions, 
the most attractive one, known as the seesaw mechanism, has been conceived 
by the authors in~\cite{YAN} within the framework of $SO(10)$ or left-right 
symmetric models. 
In these theories, right-handed neutrinos are introduced with the 
simultaneous inclusion of Majorana masses that violate the lepton-number 
($L$) by $\Delta L=2$ operators in the Yukawa sector.
The neutrino-mass spectrum of a simple seesaw model with one generation 
of quarks and leptons consists of two massive Majorana neutrinos, 
$\nu$ and $N$, having masses $m_\nu \simeq m^2_D/m_M$ and $m_N\simeq
m_M$. If the Dirac mass term $m_D$ is of the order of a typical charged-lepton
or quark mass, as dictated by GUT relations~\cite{GJ}, and the Majorana-mass
scale $m_M$ is sufficiently large, one can then obtain a very light neutrino
$\nu$. The general situation of an interfamily seesaw-type model with 
a number $\nG$ of weak isodoublets and a number
$\nR$ of right-handed neutrinos is more involved~\cite{ZPC} and will be 
discussed in Section 2. 

If nature keeps to the pathway of a seesaw-type solution, 
then heavy Majorana neutrinos at the mass scale of TeV may 
manifest themselves in $L$-violating processes at high-energy 
$ee$~\cite{PROD1ee,HM}, $ep$~\cite{PRODep}, and $pp$ 
colliders~\cite{PRODpp,PROD}, in possible lepton-flavour-violating 
decays of the $Z$~\cite{KPS} and Higgs particles ($H$)~\cite{APetal} 
or through universality-breaking effects in leptonic diagonal 
$Z$-boson decays~\cite{BKPS}. 
Their existence may also influence~\cite{BS,Hansi} the size of electroweak 
oblique parameters~\cite{STU,XYZ}, tri-gauge boson $WWZ$- and
$ZZZ$-couplings~\cite{BP}, or specific Higgs observables considered 
recently~\cite{IKP,Bernd}.
Finally, there are many other places scanned by exhaustive combined analyses 
of charged-current-universality effects in leptonic $\pi$ 
decays, neutral-current interactions in neutrino-nucleon scatterings, 
$\tau$-polarization asymmetries,
neutrino-counting experiments at the CERN Large Electron Positron Collider
(LEP), etc.~\cite{LL,BCK}, in which Majorana 
neutrinos could also manifest their presence.

Another possible solution of the neutrino-mass problem has been contemplated 
in the framework of heterotic superstring models~\cite{EW} or 
certain scenarios of $SO(10)$ models~\cite{WW}. 
The low-energy limit of such theories extend the SM field content by adding 
new left-handed and right-handed neutral isosinglets, and assuming the 
absence of $\Delta L=2$ operators in the Yukawa sector. 
In a simple one-generation scenario, one obtains three Weyl fermions from 
which one of them is completely massless to all orders of perturbation 
theory~\cite{BSVMV} and
the other two are degenerate in mass and thus form a heavy Dirac neutrino
which has a mass of the order of the isosinglet Dirac mass $M$. The Dirac
mass $M$ simply 
connects the right-handed and left-handed chiral singlets in the Yukawa
sector. This solution is particularly preferable if the light known
neutrinos turn out to be strictly massless. The model could straightforwardly
be extended to $\nG$ generations without qualitatively changing its features
regarding neutrino masses. In an $\nG$-generation model, one generally obtains
$\nG$ massless neutrinos and $\nG$ heavy Dirac neutral 
fermions~\cite{BSVMV,RV}. 
This minimal model is invariant under 
the gauge group $SU(2)_L\otimes U(1)_Y$ and possesses many attractive 
features that might be summarized in~\cite{Marek}. For example, even if the 
total lepton number is conserved, the model does generally violate the separate 
leptonic quantum numbers and can hence account for possible $L$- and/or 
$CP$-violating signals at the $Z$ peak~\cite{BSVMV} or in other high-energy 
processes~\cite{PROD2ee}. 

In this paper we carefully study the three-body decays of a charged lepton,
$l$, into other three charged leptons, which we denote hereafter as 
$l'$, $l_1$, and $\bar{l}_2$. After detailed calculations, we find 
that the decay amplitude of 
$l\to l'l_1\bar{l}_2$ depends quadratically on the mass of the heavy 
Dirac or Majorana neutrino, which violates explicitly the decoupling
theorem~\cite{AC}. Such a nondecoupling behaviour has recently been observed 
to take place in {\em three-generation} seesaw-type models, 
where the effective couplings $Hll'$~\cite{APetal} and $Zll'$~\cite{KPS,BKPS} 
show a strong quadratic dependence of the heavy neutrino mass. In past,
similar flavour-dependent nondecoupling effects have been found in the 
one-loop amplitude of the decays $Z\to b\bar{s}$~\cite{Zbs} and 
$Z\to b\bar{b}$~\cite{Zbb,Pepe}, where the top quark plays the r\^ole of 
heavy neutrinos.

Among the various decay processes, we find numerically that the decays, 
$\tau^-\to e^-e^-e^+$ and $\tau^-\to e^-\mu^-\mu^+$ [or complementary 
the decays $\tau^-\to \mu^-\mu^-\mu^+$ and $\tau^-\to \mu^- e^- e^+$], have the 
biggest opportunity to be detected at the present or future LEP data.
Furthermore, we analyze the effect of genuine Majorana-neutrino contributions 
to these decays.

The present work is organized as follows: 
In Section 2, we give a brief description of the basic low-energy 
structure of the seesaw-type models mentioned above. 
In Section~3, we discuss general constraints that should be imposed on
these models. Analytically, Section 3.1 considers possible constraints 
based on the assumption that the model should generate a sufficiently large
lepton asymmetry via the out-of-equilibrium $L$-violating decays of a heavy 
Majorana 
neutrino which can be converted later on into the observed baryon-number ($B$) 
asymmetry in the universe (BAU) due to the sphaleron interactions. 
In section~3.2, stringent constraints of possible 
non-SM mixings are derived by a global analysis of all existing 
low-energy data. Also, bounds that may be obtained by 
the non-observation of leptonic non-diagonal $Z$-boson decays at LEP are 
discussed in Section 3.3. 
In Section~4, we analytically calculate the branching ratios of the 
photonic decays of a lepton ($l$), $l\to l'\gamma$, and the three-body decay 
modes of the type $l \to l'l_1\bar{l}_2$ in the context of the models 
discussed in Section~2.
Numerical predictions and discussion of these lepton-flavour-violating 
decays are summarized in Section~5. 
We draw our conclusions in Section~6.

\setcounter{equation}{0}
\section{Theoretical models}
\indent

In this section, we will give a short description of the basic 
low-energy structure of the two most popular extensions of the 
SM that can naturally account for very light or strictly massless 
neutrinos. The field content of these models, which could also be
motivated by heterotic superstring models~\cite{EW} or certain 
$SO(10)$-GUTs~\cite{YAN,WW}, is free of anomalies~\cite{Marek}.   
These two scenarios are: (i) the interfamily seesaw-type model realized in the 
SM with right-handed neutrinos~\cite{YAN,PRODep,ZPC} and (ii) the SM with 
left-handed and right-handed neutral singlets~\cite{EW,BSVMV,RV}.

{\em (i) The SM with right-handed neutrinos.} 
In general, the interfamily seesaw-type model, being invariant under 
the SM gauge group, represents one of the most natural framework to 
predict heavy Majorana neutrinos. 
Such a model is obtained by introducing  a number $\nR$ of right-handed 
neutrinos, $\nu^0_{Ri}$, in the SM [in addition to $\nG$ left-handed ones 
$\nu^0_{Li}$] and allowing simultaneously the presence of 
$\Delta L=2$ operators.
The Yukawa sector containing the neutrino masses is then written down as
\beq 
-\cL^\nu_Y\ =\ \frac{1}{2} (\bar{\nu}^0_L,\ \bar{\nu}^{0C}_R)\
M^\nu \left( \barr{c} \nu^{0C}_L\\ \nu^0_R \earr \right)\  +\ H.c.,
\eeq 
where the $(\nG+\nR) \times (\nG+\nR)$-dimensional  neutrino-mass matrix
\beq
M^\nu\ =\ \left( \barr{cc} 0 & m_D\\ m_D^T & m_M \earr \right).
\eeq 
The matrix $M^\nu$ can always be diagonalized by a unitary matrix 
$U^\nu$ of the same dimensionality with the neutrino-mass matrix 
({\it i.e.}, $U^{\nu T}M^\nu U^\nu =\hat{M}^\nu$). 
One then gets $\nG+\nR$ physical Majorana neutrinos $n_i$ through the 
unitary transformations
\beq
\left(\barr{c} \nu^0_L \\ \nu^{0C}_R \earr \right)_i\ =\ 
\sum_{j=1}^{\snG+\snR} U^{\nu\ast}_{ij}\ n_{Lj}\quad \mbox{and}\quad
\left(\barr{c} \nu^{0C}_L \\ \nu^0_R \earr \right)_i\ =\ 
\sum_{j=1}^{\snG+\snR} U^\nu_{ij}\ n_{Rj}.
\eeq 
The first $\nG$ neutral states, $\nu_i$ ($\equiv n_i$ for $i=1,\dots,\nG$), 
are identified with the known $\nG$ light neutrinos ({\it i.e.}, $\nG=3$),
while the remaining $\nR$ mass eigenstates, $N_j$ ($\equiv n_{j+\snG}$ for 
$j=1,\dots,\nR$), represent heavy Majorana neutrinos which are novel particles 
predicted by the model. 
The quark sector of such an extension can completely be described by the SM.   

It is important to notice that the general matrix $M^\nu$ of Eq.~(2.2) 
takes the known seesaw form~\cite{YAN} in case $m_M\gg m_D$.
Nevertheless, this hierarchical scheme can drastically be relaxed in a 
two-family seesaw-type model without contradicting experimental upper bounds 
on light-neutrino masses~\cite{ZPC,HM,PRODep}. The light-heavy neutrino mixings
of such scenarios, $s^{\nu_l}_L$, can, in principle, be scaled as 
$s^{\nu_l}_L \sim m_D/m_M$ rather than $s^{\nu_l}_L \sim \sqrt{m_{\nu_l}/m_N}$ 
as usually derived in a one-family seesaw scenario~\cite{YAN,CL}.
In other words, high Dirac mass terms are allowed to be present in $M^\nu$ 
and only the ratio $m_D/m_M$ ($\sim s^{\nu_l}_L$) gets limited by a global
analysis of low-energy and LEP observables. The latter advocates our
treatment of originally considering the mixings $s^{\nu_l}_L$ and heavy 
neutrinos masses $m_{N_i}$ as free phenomenological parameters, being
subject later on to the constraints that will be discussed in Section~3. 

Adopting the conventions of Ref.~\cite{ZPC},
the interactions of the Majorana neutrinos, $n_i$, and charged
leptons, $l_i$, with the gauge bosons, $W^\pm$ and $Z$, and the unphysical
Goldstone bosons, $G^\pm$ and $G^0$ (in the Feynman-'t Hooft gauge), 
are correspondingly obtained by the Lagrangians:
\bea
\cL_{int}^W &=& -\ \frac{g_w}{\sqrt{2}} W^{-\mu}\
\sum_{i=1}^{\snG}\sum_{j=1}^{\snG+\snR} \ B_{l_ij}\
{\bar{l}}_i  {\gamma}_{\mu} \PL \ n_j \ + \ 
H.c. ,\\[0.3cm]
\cL_{int}^Z &=& -\ \frac{g_w}{4c_w}  Z^\mu\
\sum_{i,j=1}^{\snG+\snR}
\bar{n}_i \gamma_\mu \Big[ i\Im C_{ij}\ -\ \gamma_5\Re C_{ij}
\Big] n_j
\eea 
and 
\bea
\cL^{G^\mp}_{int} &=& -\ \frac{g_w}{\sqrt{2}M_W}\ G^-\
\sum_{i=1}^{\snG}\sum_{j=1}^{\snG+\snR} \ B_{l_ij}\
\bar{l}_i \Bigg[ m_{l_i}\PL\ -\ m_j\PR  \Bigg] n_j\ 
+\ H.c.,\\[0.3cm]
\cL^{G^0}_{int} &=&  \frac{ig_w}{4M_W}\ G^0\
\sum_{i,j=1}^{\snG+\snR}
\bar{n}_i \Bigg[ \gamma_5 (m_i+m_j)\Re C_{ij}
+\ i(m_j-m_i)\Im C_{ij} \Bigg] n_j .
\eea 
where $g_w$ is the weak coupling constant, $c^2_w=1-s^2_w=M^2_W/M^2_Z$, 
$\PL(\PR) = (1+(-)\gamma_5)/2$, and $m_i$ denotes all the physical
neutrino masses. 
In Eqs.~(2.4)--(2.7), $B$ and $C$ are $\nG\times (\nR +\nG)$- and 
$(\nG +\nR)\times (\nR+\nG)$-dimensional matrices, respectively, 
which are defined as
\beq
B_{l_ij}\ = \sum\limits_{k=1}^{\snG} V^l_{l_ik} U^{\nu\ast}_{kj}\quad
\mbox{and}\quad
C_{ij}\ =\ \sum\limits_{k=1}^{\snG}\ U^\nu_{ki}U^{\nu\ast}_{kj},
\eeq 
where $V^l$ is the leptonic Cabbibo-Kobayashi-Maskawa (CKM) matrix.

Note that the flavour-mixing matrices $B$ and $C$ satisfy a 
number of identities, which are derived just by using the information of
$SU(2)_L\otimes U(1)_Y$ invariance of $\cL^\nu_Y$.  
These identities, which are forced by the renormalizability of the 
interfamily seesaw-type model, can be summarized as~\cite{APetal,ZPC}
\bea
\sum\limits_{k=1}^{\snG+\snR} B_{l_1k}B_{l_2k}^{\ast} =  {\delta}_{l_1l_2},\
\sum\limits_{k=1}^{\snG +\snR} C_{ik}C^\ast_{jk} =  C_{ij}, \ 
\sum\limits_{k=1}^{\snG +\snR} B_{lk}C_{ki}  =   B_{li}, \ 
\sum\limits_{k=1}^{\snG} B_{l_ki}^{\ast}B_{l_kj}  =  C_{ij},\\
\sum\limits_{k=1}^{\snG +\snR} m_k C_{ik}C_{jk} =  0,\ \
\sum\limits_{k=1}^{\snG + \snR} m_k B_{lk}C^\ast_{ki} =  0,\ \
\sum\limits_{k=1}^{\snG +\snR} m_k B_{l_1k}B_{l_2k} =  0.\qquad\mbox{}
\eea 
Consequently, our theoretical analysis should be regarded to be 
independent of the weak-basis structure of possible neutrino-mass-matrix
ans\"atze~\cite{ansatz}. 
It is now instructive to re-express the $Z$-boson
coupling to the Majorana neutrinos, $n_i$, as follows:
\beq
\cL_{int}^Z \ = \ -\ \frac{g_w}{4c_w}  Z^\mu\
\sum_{i,j=1}^{\snG+\snR}
\bar{n}_i \gamma_\mu \Big[ C_{ij}\PL\ -\ C^\ast_{ij}\PR
\Big] n_j.
\eeq 
One can thus remark that the coupling $Zn_in_j$ is generally 
flavour non-diagonal and has both chiralities in this minimal model.

{\em (ii) The SM with left-handed and right-handed neutral singlets.} 
Another attractive scenario predicting for the light neutrinos to be 
strictly massless serves an extension of the SM, 
in which left-handed neutral singlets, $S_{Li}$, in addition to the 
right-handed neutrinos, $\nu^0_{R_i}$, are introduced. 
Furthermore, we assume that $\Delta L=2$
interactions are absent from the model, and the number of right-handed 
neutrinos, $\nR$,  equals the number of the singlet fields $S_{Li}$. 
After the spontaneous 
break-down of the SM gauge symmetry, the Yukawa sector relevant for the 
neutrino masses reads~\cite{EW,WW}
\beq 
-\cL^\nu_Y\ =\ \frac{1}{2} (\bar{\nu}^0_L,\ \bar{\nu}^{0C}_R,\ \bar{S}_L)\
\cM^\nu \left( \barr{c} \nu^{0C}_L\\ \nu^0_R \\ S^C_L \earr \right)\  +\ H.c.,
\eeq 
where the $(n_G+2\nR)\times (\nG+2\nR)$ neutrino-mass matrix is given by
\beq
\cM^\nu \ \ =\ \ \left( \barr{ccc}
0 & m_D & 0 \\
m^T_D & 0 & M^T \\
0 & M & 0 \earr \right).
\eeq 
It is easy to see that the neutrino matrix $\cM^\nu$ conserves 
the total lepton number $L$ by assigning the lepton numbers to the neutrino 
fields: $L(\nu^0_L)=L(\nu^0_R)=L(S_L)=1$. 
Since the rank of the neutrino mass matrix in Eq.~(2.13) is $2\nR$, 
$\cM^\nu$ has $\nG$ zero eigenvalues. These $\nG$
massless eigenstates should clearly describe the ordinary light
neutrinos, $\nu_e$, $\nu_\mu$ and $\nu_\tau$~\cite{EW,WW}. 
The remaining $2\nR$ Weyl fermions are degenerate in
pairs due to the fact that $L$ is conserved and so form $\nR$ heavy 
Dirac neutrinos. In general, this viable seesaw-type model can have large 
Dirac components in $\cM^\nu$ and only the ratio $m_D/M\ (\sim s^{\nu_l}_L)$
will again be constrained for $M\apprge 100$~GeV (see also our discussion in 
Section~3). 
A nice feature of the model is that the individual leptonic quantum numbers 
may be violated~\cite{BSVMV,RV,JWFV}, even if $L$ is conserved. 
The charged-current Lagrangian can be obtained by Eq.~(2.4), 
while the neutral-current interaction is given by~\cite{BSVMV}
\beq
\cL_{int}^Z \ = \ -\ \frac{g_w}{2c_w}  Z^\mu\
\sum_{i,j=1}^{\snG+\snR} 
\bar{n}_i C_{ij} \gamma_\mu \PL n_j.
\eeq 
The matrices $B$ and $C$ for this specific model obey the sum rules in 
Eq.~(2.9), but not the identities of Eq.~(2.10).

At this stage, we must comment on the difference between the Lagrangians
(2.5) and (2.14). Since Eq.~(2.5) describe Majorana neutrinos contrary to
the Lagrangian~(2.14) where the massive neutrinos are Dirac, the strength
of the $Z\bar{n}_in_i$ coupling for identical Majorana fermions is two 
times larger than the one which may naively be red off from $\cL_{int}^Z$ 
in Eqs.~(2.5) and~(2.11). 
The off-diagonal coupling $Z\bar{n}_in_j$ (with $n_i\neq n_j$) is again 
enhanced by a factor of two, since the charge-conjugate interaction 
$Z\bar{n}_jn_i$ will equally contribute to the coupling of the $Z$-boson to 
Majorana neutrinos.
In our forthcoming calculations, we have taken into account all 
these theoretical differences in treating Majorana and Dirac fields.
In fact, we find that taking formally the limit $C^{\ast}_{ij}\to 0$
but keeping $C_{ij}\neq 0$ in Eq.~(2.11) and considering 
the afore-mentioned statistical Majorana factors is sufficient to 
recover the model with additional left-handed neutral singlets.

To make life easier, we ultimately make the following reasonable assumptions: 
To a good approximation, we assume that possible novel particles related to
the above unified theories, such as leptoquarks~\cite{PS} or extra 
charged and neutral gauge bosons, $W_R^\pm$~\cite{MS} and 
$Z'$~\cite{HJLV,BCK}, are sufficiently heavy so as to decouple completely 
from the low-energy processes discussed in Sections~3.3 and~4.
For obvious reasons, possible singlet and triplet Majoron 
fields~\cite{CMP,GR,MZ,APmaj} are considered to couple very weakly to 
matter so that we can safely ignore them in our considerations. 

\setcounter{equation}{0}
\section{General constraints on the models}
\subsection{Cosmological constraints}
\indent

Unified theories based on the gauge group $SO(10)$ or 
$E_6$ can naturally accommodate right-handed neutrinos in 
addition to quarks and leptons of the SM. In such theories, 
the Majorana mass $m_M$ can directly be related to the 
$B-L$ scale of a local symmetry which is assumed to be 
spontaneously broken. It is therefore evident that the 
mass of heavy Majorana neutrinos will be determined
by the scale of $B-L$ breaking.
Moreover, out-of-equilibrium lepton-number-violating decays of 
heavy Majorana neutrinos can generate a non-zero $L$~\cite{FT} in 
the universe through the $L$-violating interactions of 
Eqs.~(2.4)--(2.7). This excess in $L$ 
can be converted into a $B$ asymmetry of the universe via the 
$(B+L)$-violating sphaleron interactions, which are in thermal 
equilibrium above the critical temperature of the electroweak phase 
transition~\cite{KRS,COSMOS}. Many studies have recently been devoted to 
constrain the $(B-L)$-violating mass scale by making use of the drastic 
out-of-equilibrium condition for the $\Delta L=2$ operators, 
and so to derive a lower mass bound on the heavy 
Majorana neutrinos~\cite{MAL,BY,CDEO,HDGR,CKO}. 
For example, in~\cite{MAL} conceivable scenarios 
predicting heavy Majorana neutrinos with $m_N=1-10$~TeV could naturally 
account for the observed BAU. Subsequently, it was argued~\cite{BY} that
the $m_N$ lower bound of $\sim 1$~TeV was considerably underestimated 
and a lower bound on $m_N > 10^5$~TeV should be imposed in a 
two-generation scenario of right-handed neutrinos with large 
interfamily mixings so as to be compatible with the existing BAU.
This would obviously imply that probing Majorana-neutrino physics
at collider energies may not be phenomenologically interesting.

The latter observation can indeed be valid in a two-generation-mixing model with
two right-handed neutrinos. 
In general, in three-generation models 
with lepton-flavour mixings, a careful inspection of chemical potentials
has shown that the stringent mass bound of heavy Majorana neutrinos mentioned
above can be weakened dramatically
and is quite model dependent~\cite{HDGR,CKO}. In particular,
it is sufficient that in equilibrium state one individual lepton number, 
{\it e.g.\ }$L_\mu$, is conserved in order to generate the BAU via the
sphaleron interactions, even if nonvanishing operators with 
$\Delta L_{l_i}\neq \Delta L_\mu$ were in thermal equilibrium~\cite{HDGR}.
The reason is that
sphalerons generally conserve the quantum numbers $B/3-L_{l_i}$~\cite{CDEO,HDGR}
and thus preserve any BAU generated by an excess, {\it e.g.}, in $L_\mu$, 
from being washed out. Similar conclusions have been drawn in Ref.~\cite{CKO}.

A viable scenario of heavy Majorana neutrinos with masses $\sim$ TeV
can easily be realized in the SM with $\nR=4$. If the BAU is to be 
generated through an excess in muonic number density,
this asymmetry in $L_\mu$ can be achieved by considering 
a neutrino mass matrix, $M^\nu$, similar to the $CP$-violating scenario
given in~\cite{IKP}. This scenario contains one left-handed
neutrino, $\nu^0_{L1}$, to which, for the case at hand, a muonic quantum number 
should be assigned, and two right-handed neutrinos, denoted as $\nu^0_{R3}$ and 
$\nu^0_{R4}$. 
The explicit form of $M^\nu$ is then given by~\cite{IKP}
\beq
M^\nu\ \ =\ \ \left( \barr{ccc} 0 & a & b \\ a & A & 0 \\ b & 0 & B \earr
\right),
\eeq 
where $a$ and $b$ are in general complex numbers, and $A$ and $B$ can be
chosen to be real.
Out-of-equilibrium conditions for generating a sufficiently large asymmetry 
in $L_\mu$, which can give rise to the established BAU, lead to the 
stringent lower bounds on the masses of the corresponding physical 
heavy neutrinos $N_{3,4}$ as consistently obtained by~\cite{BY}. 
However, the remaining $e$- and $\tau$-lepton families can strongly mix 
each other via two {\em additional} right-handed neutrinos, {\it e.g.}, 
$\nu^0_{R1}$ and $\nu^0_{R2}$, and form an {\em individual} $4\times 4$ 
seesaw-type matrix. Operators $\Delta L_e\neq 0$ and $\Delta L_\tau\neq 0$ are 
now allowed to be in thermal equilibrium provided that $\Delta (L_e-L_\mu)=0$ 
and $\Delta (L_\tau-L_\mu)=0$. 
The latter condition is automatically satisfied due to the construction
of this specific scenario with $\nR=4$. As a consequence, the severe lower 
mass bounds on the physical heavy neutrinos $N_1$ and $N_2$ can be 
evaded completely. A similar analysis in a SM with $\nR=3$ is more involved 
due to the flavour-mixing effects in the neutrino-mass matrix and can be given 
elsewhere. 

\subsection{Low-energy constraints}
\indent

There exists a great number of low-energy experiments that could set upper 
bounds on possible non-SM couplings~\cite{LL}. The most significant 
experimental tests giving stringent constraints turn out to be 
the neutrino counting at the $Z$ peak, the precise measurement of the 
muon width $\mu \to e\nu_e\nu_\mu$, charged-current universality effects on the 
observable $\Gamma(\pi \to e\nu)/\Gamma(\pi\to \mu\nu)$, non-universality
effects on
$B(\tau \to e \nu \nu)/B(\tau \to \mu \nu\nu)$, and other nuclear
physics effects and experiments. All these constraints, which are derived 
by the low-energy data mentioned above, depend crucially on the gauge structure
of the model under discussion. For example, assuming the supersymmetric (SUSY) 
nature of the $E_6$ models or SUSY-$SO(10)$ unified theories~\cite{RJNP,GGRGR},
and $R$-parity invariance, the neutralino state could then be the lightest 
supersymmetric particle (LSP) which is stable. If the mass of the LSP is 
assumed to be in the vicinity of $M_Z/2$, then an additional invisible decay 
channel for the $Z$ boson will open kinematically and neutrino-counting limits 
imposed on the couplings $Z\nu_i\nu_j$ may not be applicable. 
Furthermore, an analysis of decays of the type $Z\to N\nu$,
which have been considered in~\cite{PROD2ee}, suggests that $m_N \apprge 
100$~GeV for $(s^{\nu_l}_L)^2\sim 0.01$. 

Thus, identifying the non-SM-mixing angles $(s^{\nu_l}_L)^2$ of Ref.~\cite{LL} 
as
\beq
(s^{\nu_l}_L)^2 \equiv \sum\limits_{j=1}^{\snR} |B_{lN_j}|^2,
\eeq 
and in view of the discussion given above, one may tolerate the 
following upper limits~\cite{LL}:
\beq
(s^{\nu_e}_L)^2\ \ < \ \ 0.015, \quad 
(s^{\nu_\tau}_L)^2\ \ < \ \ 0.070,\quad \mbox{and}\quad 
(s^{\nu_\mu}_L)^2\ \ < \ \ 1.\ 10^{-9}.
\eeq 
In Eq.~(3.3), the tight upper bound on $s^{\nu_\mu}_L$ represents 
that the muonic quantum number is practically conserved in thermal
equilibrium. Note that, without loss of generality, one could equally 
interchange the upper limit on $(s^{\nu_e}_L)^2$ with that of 
$(s^{\nu_\mu}_L)^2$. To be precise, we will assume 
$(s^{\nu_\mu}_L)^2\simeq 0$, in what follows.

Another limitation to the parameters of our model comes from the 
requirement of the validity of perturbative unitarity that can be violated
in the limit of large heavy-neutrino masses.
A qualitative estimate for the latter may be obtained by requiring that 
the total widths, $\Gamma_{N_i}$, and masses of neutrino fields $N_i$
satisfy the inequality 
\beq
\frac{\Gamma_{N_i}}{m_{N_i}}\ \ < \ \ \frac{1}{2}. 
\eeq 
The total widths of the heavy neutrinos, $\Gamma_{N_i}$,
can be written down as a sum over all 
possible decay channels~\cite{ZPC}, {\it i.e.}
\beq
\Gamma_{N_i} = \sum_{l_j} \Gamma (N_i\to l_j^\pm W^\mp)\ +\
\sum_{\nu_j} \Big( \Gamma (N_i\to \nu_j Z)\ +\ \Gamma (N_i\to \nu_j H)\Big).
\eeq 
In the limit of $m_{N_i} \gg M_W,\ M_Z,\ M_H$, the above expression 
simplifies to
\beq
\Gamma_{N_i}\ \ =\ \ \frac{\alpha_w}{4 M^2_W}\ m^3_{N_i}\ |C_{N_iN_i}|^2,
\eeq 
with $\alpha_w=g^2_w/4\pi$.

\subsection{Constraints from leptonic {\boldmath $Z$}-boson decays}
\indent

Aside from low-energy constraints discussed in Section~3.2,
many extensions of the SM derived by unified theories may give rise to 
lepton-flavour-violating decays of the $Z$ boson~\cite{LFstring,LFCNC,KPS}.
In particular, it has been found in~\cite{KPS} that the non-observation of 
such non-SM signals at LEP may impose combined bounds both on heavy 
neutrino masses $m_{N_i}$ and mixings $(s^{\nu_l}_L)^2$. 
The reason is that the amplitude of such a decay depends quadratically on
the heavy neutrino mass, leading to measurable rates.
In a self-explanatory way, the amplitude of the decay 
$Z\to ll'$ may generally be parametrized as 
\beq     
\cT(Z\rightarrow \bar{l} l')\ \ =\ \ \frac{ig_w\alpha_w}{8\pi c_w}\
    \cF_Z^{ll'}\ \veps_Z^\mu \ \bar{u}_{l'}\gmd(1-\g5)v_l,
    \eeq 
where $\alpha_w=g_w^2/4\pi$ and the form factor $\cF^{ll'}_Z$, which is 
induced by the Feynman graphs of Fig.~2 at the one-loop electroweak order, 
is given in Appendix A. The branching ratio of this decay mode is
obtained by
\beq     
B(Z\rightarrow \bar{l} l'+\bar{l}'l)\ \ =\ \ 
\frac{\alpha_w^3}{48\pi^2c_w^3}\frac{M_W}{\Gamma_Z}
                              \mid \cF_Z^{ll'} \mid^2, 
\eeq 
where $\Gamma_Z=2.49$~GeV is the experimental value of 
the total width of the $Z$ boson~\cite{PDG}. 
We postpone the numerical discussion of possible constraints that might
arise due to lepton-flavour-violating decays of the $Z$ boson in Section~5.
 
\setcounter{equation}{0}
\section{Flavour-violating decays of charged leptons}
\indent

In Sections~4.1 and~4.2, we will theoretically analyze the possibility 
of lepton flavour nonconservation in decays of the form $l\to l'\gamma$ and 
$l\to l'l_1\bar{l}_2$, respectively. As mentioned in the Introduction,
$l$, $l'$, $l_1$ and $l_2$ denote usual charged leptons, {\it i.e.\ }the 
$e$, $\mu$ and $\tau$ leptons.

\subsection{The decay {\boldmath $l\to l'\gamma$} }
\indent

In the framework of the minimal class of models discussed in Section 2, 
heavy Majorana or Dirac neutrinos can give rise to the decay $l\to l'\gamma$.
The Feynman graphs responsible for such a decay are shown in Fig.~1. 
Applying electromagnetic gauge invariance to the decay amplitude 
$l(p)\to l'(p') \gamma (q)$, where the photon, $\gamma$, 
may be off-mass shell, yields~\cite{Cheng,IL}
\bea
  \cT(l\rightarrow l'\gamma)&=&-i\frac{e\alpha_w}{16\pi M_W^2}
   \veps_\gamma^\mu\
     \bar{u}_{l'}\Bigg[ \sum_{i=1}^{\snG +\snR} B^\ast_{li} B_{l'i}
         F_\gamma(\li)(q^2\gmd-\qsl\qmd)(1-\g5)
    \nonumber\\
& &  - \sum_{i=1}^{\snG +\snR} B^\ast_{li}B_{l'i}
         G_\gamma(\li)\;i\smnd\qnu\Big( m_{l'}(1-\g5)+m_l(1+\g5)\Big)\Bigg] u_l,
\eea 
where $\li=m^2_i/M^2_W$, $q=p-p'$ denotes the outgoing momentum of the photon, 
and the form factors $F_\gamma$ and $G_\gamma$ are given in Appendix B. It is 
now straightforward to calculate the branching ratio of $l\to l'\gamma$
\beq
  B(l\rightarrow l'\gamma)\ \ =\ \ \frac{\alpha^3_w s_w^2}{256\pi^2}
                            \frac{m_l^4}{M_W^4}\frac{m_l}{\Gamma_l}
                            \mid G_\gamma^{ll'}\mid^2,
\eeq 
where $\Gamma_l$ is the total width of the decaying lepton $l$, while
$G_\gamma^{ll'}$ in Eq.~(4.2) represents a composite form factor defined 
in Appendix B. Specifically, for the total width of the $\tau$ lepton, 
we use the experimental value, $\Gamma_\tau=2.1581\ 10^{-12}$~GeV~\cite{PDG},
whereas the muon total width may be obtained by~\cite{Beg} 
\beq
\Gamma_\mu=\frac{G_F^2 m_\mu^5}{192\pi^3}\left(1-8\frac{m_e^2}{m_\mu^2}\right)
\left[ 1+\frac{\alpha_{em}}{2\pi}\Big( \frac{25}{4}-\pi^2\Big)\right],
\eeq 
where $\alpha_{em}=e^2/4\pi$. The muon total width given in Eq.~(4.3) is 
in excellent agreement with the experimental value reported in~\cite{PDG}.

The experimental upper bounds arising from the non-observation of 
decays of the type $l\to l'\gamma$ are~\cite{PDG}
\beq
B(\tau\to e\gamma)< 1.2\ 10^{-4},\quad B(\tau \to \mu\gamma) <4.2\ 10^{-6},
\quad B(\mu\to e \gamma) < 4.9\ 10^{-11},
\eeq 
at $90\%$~CL. Using the values for the mixing angles $(s^{\nu_l}_L)^2$
of Eq.~(3.3), one easily finds that the photonic decays involving 
muons are extremely suppressed in our minimal scenarios. Furthermore,
the theoretical prediction $B(\tau \to e\gamma)\apprle 10^{-7}$ 
shows that photonic decays of a $\tau$ lepton may not be the most favourable 
place to probe heavy neutrino physics. 
 
\subsection{Three-body leptonic decays {\boldmath
$l\to l' l_1 \bar{l}_2$ } }
\indent

In a three-generation model the decaying charged lepton $l$ will either
be a muon or a $\tau$ lepton. There are seven possible decays of the 
generic form $l\to l'l_1\bar{l}_2$  
\bea
 &\mbox{a}.& \ \tau^-\to \mu^-\mu^-\mu^+, \nonumber\\
 &\mbox{b}.& \ \tau^-\to \mu^-\mu^-e^+ ,  \nonumber\\
 &\mbox{c}.& \ \tau^-\to e^-\mu^-\mu^+ ,  \nonumber\\
 &\mbox{d}.& \ \tau^-\to e^- e^-\mu^+  ,  \nonumber\\
 &\mbox{e}.& \ \tau^-\to e^-\mu^- e^+  ,   \nonumber\\
 &\mbox{f}.& \ \tau^- \to e^- e^- e^+  ,   \nonumber\\
 &\mbox{g}.& \ \mu^-\to e^- e^- e^+    .
\eea 
To facilitate our computational task, we divide the decays in Eq.~(4.5) 
into three categories according to the leptonic flavours in the final 
state: Category (i) contains all the decays where $l'\neq l_2$ and 
$l_1=l_2$ or $l'=l_2$ and $l_1\neq l_2$ ({\it i.e.} the decays (c) and (e)). 
Category (ii) comprises all the decays where $l'=l_1=l_2$
({\it i.e.} the decays (a), (f) and (g)). And lastly,
all the decays with final leptons having $l'\neq l_2$, $l_1\neq l_2$
belong to the category (iii) ({\it i.e.} the decays (b) and (d)). 

The transition amplitude of the decay $l(p)\to l'(p')l_1(p_1)\bar{l}_2(p_2)$ 
receives contributions from $\gamma$- and $Z$-mediated graphs shown in Figs.~1 
and 2, respectively, and box diagrams given in Fig.~3. These three different
amplitudes are conveniently written down as follows: 
\bea
  \cT_\gamma(l\rightarrow l' l_1 \bar{l}_2)&=&
     -\frac{i\alpha_w^2s_w^2}{4M_W^2}
     \delta_{l_1 l_2}\sum_{i=1}^{\snG+\snR} 
    B^\ast_{li}B_{l'i}\ \bar{u}_{l_1}\gmu v_{l_2}\ 
   \bar{u}_{l'}\Big[ F_\gamma(\li)(\gmd-\frac{\qmd\qsl}{q^2})(1-\g5)
    \nonumber\\
& &              -iG_\gamma(\li)\smnd\frac{\qnu}{q^2}
                                (m_l(1+\g5)+m_{l'}(1-\g5))\Big] u_l,
     \\
& & \nonumber\\
  \cT_Z(l\rightarrow l' l_1 \bar{l}_2)&=&-\frac{i\alpha_w^2}{16M_W^2}\
     \bar{u}_{l'}\gmd(1-\g5)u_l\ \bar{u}_{l_1}\gmu(1-4s_w^2-\g5)v_{l_2}
    \nonumber\\
& & \times  \delta_{l_1 l_2}\sum_{i,j=1}^{\snG+\snR} B^\ast_{li}B_{l'j}
 \Big[ \delta_{ij}F_Z(\li)+C_{ij}H_Z(\li,\lj)+C^\ast_{ij} G_Z(\li,\lj)\Big],\
    \ \, \\
& & \nonumber\\
  \cT_{Box}(l\rightarrow l' l_1 \bar{l}_2)&=&-\frac{i\alpha_w^2}{16M_W^2}\
     \bar{u}_{l'}\gmd(1-\g5)u_l\ \bar{u}_{l_1}\gmu(1-\g5)v_{l_2} 
    \nonumber\\
& &\times \sum_{i,j=1}^{\snG +\snR} 
    \Big[ (B_{l'i}B_{l_1j}+B_{l_1i}B_{l'j})B^\ast_{li} B^\ast_{l_2j}
    \: F_{Box}(\li,\lj) 
    \nonumber\\
& & +B_{l'i}B_{l_1i}B^\ast_{lj} B^\ast_{l_2j}\: G_{Box}(\li,\lj) \Big],
\eea 
where $q=p_1+p_2$. In addition to the photonic form factors $F_\gamma$ and 
$G_\gamma$ in Eq.~(4.6), the form factors $F_Z$, $H_Z$, $G_Z$, $F_{Box}$,
and $G_{Box}$ are given in Appendix B. Note that the term proportional
to $G_\gamma$ in Eq.~(4.6) contains a non-local interaction which is singular
in the limit $q^2\to 0$.

Following the classification mentioned above, the branching ratio
for all decays belonging to the first category is found to be
\footnote[1]{In our calculations we have used a notation similar 
to the authors in~\cite{JWFV}. However, our branching-ratio expressions~(4.9) 
and~(4.10) are at variance with their results.}
\bea
\lefteqn{ B(l^-\rightarrow l'^- l_1^- l_2^+, l'\neq l_2, l_1=l_2)\ \ =\ \ 
     \frac{\alpha_w^4}{24576\pi^3}\ \frac{m_l^4}{M_W^4}\
     \frac{m_l}{\Gamma_l} }\hspace{1cm}
     \nonumber\\
& &  \times \Big\{ \mid F_{Box}^{l l' l_1 l_1}+F_Z^{l l'}
            -2s_w^2(F_Z^{l l'}-F_\gamma^{l l'})\mid ^2 \,
     +\ 4s_w^4\mid F_Z^{l l'}-F_\gamma^{l l'}\mid ^2 
     \nonumber\\
& &  +\ 8s_w^2\: \Re [(F_Z^{l l'}+F_{Box}^{l l' l_1 l_1})G_\gamma^{l l'\:\ast}]
     \ -\ 32s_w^4\: \Re [(F_Z^{l l'}-F_\gamma^{l l'})G_\gamma^{l l'\:\ast}] 
     \nonumber\\
& &  +\ 32s_w^4\mid G_\gamma^{l l'}\mid^2\:
     [\ln\frac{m_l^2}{m_{l_1}^2}-3]\Big\}, 
\eea 
where $F_\gamma^{ll'}$, $G_\gamma^{ll'}$, $F_Z^{ll'}$, and
$F_{Box}^{ll'l_1l_2}$ are composite form factors defined explicitly 
in Appendix B.
The branching ratios referring to the categories (ii) and (iii)
are correspondingly given by
\bea
\lefteqn{  B(l^-\rightarrow l'^- l_1^- l_2^+, l'=l_1=l_2)\ \ =\ \ 
     \frac{\alpha_w^4}{24576\pi^3}\ \frac{m_l^4}{M_W^4}\
     \frac{m_l}{\Gamma_l} }\hspace{1cm}
     \nonumber\\
& &  \times\Big\{ 2\mid {\textstyle \frac{1}{2}}
        F_{Box}^{l l_1 l_1 l_1}+F_Z^{l l_1}
            -2s_w^2(F_Z^{l l_1}-F_\gamma^{l l_1})\mid ^2
     +\ 4s_w^4\mid F_Z^{l l_1}-F_\gamma^{l l_1}\mid ^2 
     \nonumber\\
& &  +\ 16s_w^2\: \Re [(F_Z^{ll_1}+{\textstyle \frac{1}{2}}
     F_{Box}^{l l_1 l_1 l_1})
      G_\gamma^{l l_1\:\ast}]
     \ -\ 48s_w^4\: \Re [(F_Z^{ll_1}-F_\gamma^{ll_1})G_\gamma^{ll_1\:\ast}] 
     \nonumber\\
& &  +\ 32s_w^4\mid G_\gamma^{ll_1} \mid^2\: 
     [\ln\frac{m_l^2}{m_{l_1}^2}-\frac{11}{4}]\Big\}
\eea 
and
\bea
 B(l^-\rightarrow l'^- l_1^- l_2^+, l_1\neq l_2, l'\neq l_2)\ \ =\ \ 
     \frac{\alpha_w^4}{49152\pi^3}\ \frac{m_l^4}{M_W^4}\
     \frac{m_l}{\Gamma_l}\ \mid F_{Box}^{l l' l_1 l_2}\mid^2.
\eea 
Equations~(4.9) and~(4.10) contain a non-local interaction in terms
$\propto G^{ll'}_\gamma$ and $G^{ll_1}_\gamma$, which is discussed
in detail in Appendix C.
In Eq.~(4.10), one has to take into account  statistical symmetrization 
factors for the two identical final leptons ({\it i.e.}~$l'=l_1$), as well as 
additional Feynman graphs resulting from the interchange of the lepton
$l'$ with $l_1$. The set of decays in (iii) can only be induced by the
box graphs shown in Fig.~3. The amplitude of such a class of decays 
({\it i.e.}~decays (b) and (d)) is
always proportional to $(s^{\nu_l}_L)^2s^{\nu_e}_Ls^{\nu_\mu}_L$ and the 
corresponding branching ratios are hence expected to be vanishingly small 
even if one uses the upper value of $s^{\nu_\mu}_L$ in Eq.~(3.3). 
For reasons of mere academic interest, we simply note that 
$B(\tau^- \to e^- e^- \mu^+), B(\tau^-\to \mu^-\mu^- e^+) \apprle 10^{-12}$.
As a consequence, we find that the decays (c) and (f) in Eq.~(4.5)
deserve the biggest attention and will hence be analyzed numerically in the next
section.

\setcounter{equation}{0}
\section{Numerical evaluation and discussion}
\indent

We will now investigate the phenomenological impact of the two seesaw-type
models outlined in Section~2.  
In order to pin down numerical predictions, we will, for definiteness, 
assume an extension of the SM by two right-handed neutrinos. 
The neutrino mass spectrum of such a model 
consists of three light Majorana neutrinos which have been identified 
with the three known neutrinos, $\nu_e$, $\nu_\mu$, and $\nu_\tau$, and 
two heavy ones denoted by $N_1$ and $N_2$. As already mentioned in Section
2, the seesaw-type extension of the SM with one left-handed and one 
right-handed chiral singlets can effectively be recovered by the SM with 
two right-handed neutrinos when taking the degenerate mass limit for the 
two heavy Majorana neutrinos. 
It is therefore obvious that branching-ratio results for the SM with
one left-handed and one right-handed neutral singlets can be red off
from the SM with two right-handed neutrinos in the specific case 
$m_{N_1}=m_{N_2}=m_N$.

Apart from the two heavy Majorana neutrino masses which are free parameters
of the theory, the model contains numerous mixing angles, $B_{li}$
and $C_{ij}$, for which the only restriction comes from a low-energy analysis
as discussed in Sections~3.2 and~3.3. However, in our minimal model with two
right-handed neutrinos one can derive, with the help of the identities
in Eqs.~(2.9) and~(2.10), the useful relations
\beq
B_{lN_1}\ =\ \frac{\rho^{1/4} s^{\nu_l}_L}{\sqrt{1+\rho^{1/2}}}\ , \qquad
B_{lN_2}\ =\ \frac{i s^{\nu_l}_L}{\sqrt{1+\rho^{1/2}}}\ ,
\eeq 
where $\rho=m^2_{N_2}/m^2_{N_1}$. The mixings
$C_{N_iN_j}$ can also be obtained by employing Eq.~(2.9). In this way,
one gets
\bea
C_{N_1N_1} &=& \frac{\rho^{1/2}}{1+\rho^{1/2}}\ \sum\limits_{l=1}^{\snG} 
(s^{\nu_l}_L)^2, \qquad C_{N_2N_2}\ =\ \frac{1}{1+\rho^{1/2}}\
\sum\limits_{l=1}^{\snG} (s^{\nu_l}_L)^2, \nonumber\\
C_{N_1N_2}&=& -C_{N_2N_1}\ =\ \frac{i\rho^{1/4}}{1+\rho^{1/2}}\
\sum\limits_{l=1}^{\snG} (s^{\nu_l}_L)^2.
\eea 
Evidently, our minimal scenario depends only on the masses of the heavy 
Majorana neutrinos, $m_{N_1}$ and $m_{N_2}$ (or equivalently on 
$m_{N_1}$ and $\rho$), and the mixing angles $(s^{\nu_l}_L)^2$,
which are directly constrained by a global analysis of low-energy 
data.

In our illustrative model, with the help of Eq.~(5.2), we can easily obtain 
the maximal heavy neutrino mass allowed by perturbative unitarity.
Satisfying Eq.~(3.4) for both heavy neutrinos $N_1$ and $N_2$, one
gets the global relation
\beq
m^2_{N_1} \leq \frac{2M^2_W}{\alpha_w}\ \frac{1+\rho^{-1/2}}{\rho^{1/2}}
\left[\ \sum_{l=1}^{\snG} (s^{\nu_l}_L)^2\ \right]^{-1},
\eeq 
with $\rho \geq 1$. Condition~(5.3) has thoroughly been used in our 
numerical estimates to impose an upper bound on $m_{N_{1,2}}$.

For reasons mentioned in Section 4.2, we present the branching 
ratios for the leptonic decays $\tau^- \to e^- e^- e^+$ and 
$\tau^- \to e^- \mu^- \mu^+$ in Fig.~4. To gauge to which extend 
our minimal model can predict measurable rates, we have first assumed the 
maximally allowed values~\cite{LL} for $(s^{\nu_\tau}_L)^2=0.07$ and 
$(s^{\nu_e}_L)^2=0.015$ $((s^{\nu_\mu}_L)^2\simeq 0)$ given in Eq.~(3.3).
From Fig.~4 we find the encouraging branching ratios
\beq
B(\tau^- \to e^- e^- e^+) \ \apprle\ 2.\ 10^{-6}\quad \mbox{and}
\quad
B(\tau^- \to e^- \mu^- \mu^+)\ \apprle\ 1.\ 10^{-6}.
\eeq 
The present experimental upper limits on these decays are given by~\cite{PDG}
\beq
B(\tau^- \to e^- e^- e^+)\ <\ 1.3\ 10^{-5}\quad \mbox{and}\quad
B(\tau^- \to e^- \mu^- \mu^+) \ < \  1.9\ 10^{-5},\quad \mbox{CL}=90\%.
\eeq 
Even if we assume smaller values for the mixing angles, 
$(s^{\nu_\tau}_L)^2=0.035$ and $(s^{\nu_e}_L)^2=0.01$ $((s^{\nu_\mu}_L)^2=0)$, 
the lepton-flavour-violating decays of the $\tau$ lepton can still be 
significant. From Fig.~5 one has that 
\beq
B(\tau^- \to e^- e^- e^+) \ \apprle\ 5.\ 10^{-7}\quad \mbox{and}
\quad
B(\tau^- \to e^- \mu^- \mu^+)\ \apprle\ 3.\ 10^{-7},
\eeq 
and the possibility of observing such decays at future LEP experiments 
appears feasible. 
Note that the branching ratio increases with the heavy neutrino mass
to the fourth power and hence allows to reach measurable values. 
To demonstrate this fact, we have just neglected contributions of 
seemingly suppressed terms $\cO ((s^{\nu_l}_L)^4) $ in the transition 
elements and found a reduction
of our numerical values up to $\sim 10^{-2}$. In the low-mass range of
heavy neutrinos ({\it i.e.}~for $m_{N_i}< 200$~GeV) the difference between 
the two distinct computations is quite small and consistent with results 
obtained in~\cite{JWFV}. In the high-mass regime, however,
the situation changes drastically (see also Figs.~4 and 5), since in
the transition amplitude, terms proportional
to $(s^{\nu_l}_L)^2$ depend logarithmically on the heavy neutrino
mass $m_N$, {\it i.e.}~$\ln(m_N^2/M^2_W)$, while terms of 
$\cO ((s^{\nu_l}_L)^4)$ show a strong quadratic dependence in the heavy 
neutrino mass, {\it i.e.}~$m_N^2/M^2_W$. 

Fig.~6 represents genuine Majorana-neutrino quantum effects, since we have 
computed the branching ratios as a function of the ratio $m_{N_2}/m_{N_1}$
for the selective values of $m_{N_1}=200$~GeV and 500~GeV.
Although the most stringent constraints on the heavy 
Majorana neutrino masses result from Eq.~(5.3), it is, however, important
to notice that for lower neutrino masses the maximum of 
$B(\tau^-\to e^-e^-e^+)$ and $B(\tau^-\to e^-\mu^-\mu^+)$ is not given by the 
degenerate case where $\rho=1$. In fact, if $m_{N_2}/m_{N_1}\simeq 3$ the 
branching ratios show up a maximum which can be up to two times bigger than 
the case where both the heavy neutrinos, $N_1$ and $N_2$, are degenerate. 
We have thus found that for $m_{N_1}=500$~GeV and $m_{N_2}\simeq 1.5$~TeV,
\beq
B(\tau^- \to e^- e^- e^+)\ \apprle\ 2.\ 10^{-8}\quad \mbox{and}
\quad
B(\tau^- \to e^- \mu^- \mu^+)\ \apprle\ 1.5\ 10^{-8}.
\eeq 
Such effects might be accessible at $\tau$ factories if one assumes 
an upgrade in the luminosity of the LEP collider by a factor of 10.

In the following, we will discuss possible constraints that 
might arise from lepton-flavour-violating decays of the $Z$ boson.
Since we always assume that $(s^{\nu_\mu}_L)^2=0$, we will focus our
analysis on the decays $Z\to e^-\tau^+ + e^+\tau^-$. Within the perturbatively
allowed range of heavy neutrino masses as determined by Eq.~(5.3), Fig.~7 gives
\bea
B(Z\to e^-\tau^+\ +\ e^+\tau^-)\ \apprle \ 4.0\ 10^{-6}, \quad\mbox{for}
\quad (s^{\nu_\tau}_L)^2=0.070, \quad (s^{\nu_e}_L)^2=0.015,\nonumber\\
B(Z\to e^-\tau^+\ +\ e^+\tau^-)\ \apprle \ 1.1\ 10^{-6}, \quad\mbox{for}
\quad (s^{\nu_\tau}_L)^2=0.035, \quad (s^{\nu_e}_L)^2=0.010,\nonumber\\
B(Z\to e^-\tau^+\ +\ e^+\tau^-)\ \apprle \ 6.0\ 10^{-7}, \quad\mbox{for}
\quad (s^{\nu_\tau}_L)^2=0.020, \quad (s^{\nu_e}_L)^2=0.010.
\eea 
In our numerical estimates, we have used values for $(s^{\nu_l}_L)^2$
compatible with the updated upper bounds given in Eq.~(3.3).
Although all the branching ratios in Eq.~(5.8) could be detected at 
future LEP data, they cannot impose any severe constraints on 
the $\tau $ decays into three charged leptons for the present analysis. 
The experimental sensitivity at LEP is currently given by~\cite{PDG}
\beq
B(Z\to e^-\tau^+\ +\ e^+\tau^-)\ \ <\ \ 1.3\ 10^{-5},\quad \mbox{CL}=95\%.
\eeq 
Here, some comments are in order.
In Fig.~7, the branching ratios for the three different mixing-angle sets 
in the order stated in Eq.~(5.8) 
show a minimum at the positions $m_N=700$, 900 and 1200~GeV, respectively. 
The reason is that $\cO ((s^{\nu_l}_L)^2)$ and $\cO ((s^{\nu_l}_L)^4)$
terms of $\cF^{ll'}_Z$ in Eq.~(3.7) cancel each other and the whole 
transition amplitude becomes pure absorptive.
In the range of very heavy neutrinos, terms proportional to $(s^{\nu_l}_L)^4$ 
will dominate in the amplitude for the same reasons mentioned above.
The effect of such a dynamical cancellation of the dispersive part
of the amplitude could be shown up as a difference between the charge-conjugate
decay modes of $Z\to e^-\tau^+$ and $Z\to e^+\tau^-$, leading to sizeable
$CP$-violating effects~\cite{RV}.  

In Fig.~8, we display genuine Majorana-neutrino virtual effects
by examining the behaviour of the branching ratio as a function 
of the quantity $m_{N_2}/m_{N_1}$ for rather modest values of 
$m_{N_1}$. Here, the situation is more involved and depends strongly 
on the value of $m_{N_1}$ we choose. The fact that the amplitude could be 
dominated by $(s^{\nu_l}_L)^2$ terms for relatively light heavy Majorana 
neutrinos ({\it i.e.}~$m_{N_1} < 400$~GeV) or by $(s^{\nu_l}_L)^4$ terms
for larger values of $m_{N_1}$ plays a crucial r\^ole for the shape of 
the different lines drawn in Fig.~8. The common feature is, however, that
the case where both the heavy Majorana neutrinos, $N_1$ and $N_2$, have 
the same mass does not again correspond to the situation yielding 
the biggest branching-ratio value.

Finally, $\tau$ leptons can also decay hadronically via the channels:
$\tau \to l_i\eta$, $\tau\to l_i\pi^0$, etc.~\cite{JWFV}.
However, the present experimental sensitivity to these decays seems to be
rather weak~\cite{PDG} so as to set constraints on our analysis. For example, 
$B(\tau\to e\pi^0)<1.4\ 10^{-4}$, at CL$=90\%$. 

\section{Conclusions}
\indent 

We have explicitly shown that seesaw-type extensions of the minimal SM, 
which naturally contain left-handed and/or right-handed weak isosinglets, 
can favourably account for sizeable branching ratios of $\tau$ decays 
into three charged leptons that can be as large as $\sim 10^{-6}$. 
Using updated constraints for the mixings $(s^{\nu_l}_L)^2$, we have found 
that our numerical estimates of $B(\tau\to eee)$ and $B(\tau \to e\mu\mu)$ 
are in qualitative agreement with those obtained in~\cite{JWFV} for 
$m_N<200$~GeV. 
However, our branching-ratio values can be up to 100 times larger than 
the results reported in~\cite{JWFV} when the heavy neutrinos have 
TeV~masses. The reason is that the flavour-violating decays of the 
$\tau$ lepton show a strong {\em quadric} mass dependence of the heavy 
neutrino mass in a complete calculation, which gives a unique chance for 
such decays to be seen at LEP or planned collider machines.

Apart from general constraints that our minimal models should satisfy
and have been taken into account, we have found that 
$B( Z\to e^-\tau^+ + e^+\tau^-) \apprle 4.\ 10^{-6}$ within the range 
allowed by perturbative unitarity. The latter do not yet impose any 
stringent constraints on the phenomenological parameters of the theory. 
Moreover, heavy Majorana neutrinos introduce a different 
behaviour in the transition amplitude via loop effects 
as compared to heavy Dirac ones.
For example, Fig.~6 shows that an appreciably large mass 
difference between the two heavy Majorana neutrinos $N_1$ and $N_2$
({\it i.e.}~$m_{N_2}/m_{N_1} \simeq 3$) will give rise to an enhancement of a 
factor of two to the corresponding branching-ratio value obtained for 
$m_{N_1}=m_{N_2}$. 

\vskip2cm
\noindent
{\bf Acknowledgements.} We thank 
S.A.\ Abel, G.G.\ Ross, and M.\ Shaposhnikov 
for discussions about cosmological constraints on models with 
Majorana neutrinos, R.J.N.~Phillips for useful comments,  
B.A.~Kniehl for technical details of loop integrals, and
M.C.~Gonzalez-Garcia and J.W.F.~Valle for remarks and comments.

\newpage
\setcounter{section}{0}
\def\theequation{\Alph{section}.\arabic{equation}}
\begin{appendix}
\setcounter{equation}{0}
\section{Loop integrals of leptonic {\boldmath $Z$}-boson decays}
\indent

After computing the Feynman graphs shown in Fig.~2, we find that the 
analytic expression of the form factor $\cF_Z^{ll'}$ defined in Eq.~(3.7) 
can be cast into the form~\cite{KPS}
\bea
\cF_Z^{ll'}&=&\hspace{-1pt}
           \sum_{i,j=1}^{\snG+\snR} B_{l'i}B^\ast_{lj}\Bigg\{ \delta_{ij}
           \Bigg[ -\tilde{I}(\li)-3c_W^2 L_1(\li)
                 -s_W^2\li I(\li)
\nonumber\\
&&        \hspace{-8pt}
                  -\frac{1}{8}(1-2s_W^2)\li\left( 2L_1(\li)+\frac{3}{2}
                  -\frac{3}{1-\li}-\frac{(\li+2)\li\,\ln\li}{(1-\li)^2}\right)
           \Bigg]
\nonumber\\
&&          \hspace{-8pt}
             +C_{ij}\left( \frac{1}{2}L_2(\li,\lj)-\frac{1}{2}\lZ
             [K_1(\li,\lj)-K_2(\li,\lj)+\tilde{K}(\li,\lj)]
             -\frac{1}{4}\li\lj K_1(\li,\lj)\right)
\nonumber\\
&&          \hspace{-8pt}
             +C^\ast_{ij}\sqrt{\li\lj}\left( \frac{1}{2} K_1(\li,\lj)
             +\frac{1}{4} \lZ\tilde{K}(\li,\lj)-
                                 \frac{1}{4} L_2(\li,\lj)\right)\Bigg\},
\eea 
where $\li=m^2_i/M^2_W$, $\lZ=M^2_Z/M^2_W$, and
the definition of the loop integrals, $I$, $\tilde{I}$, $L_1$,
$K_1$, $K_2$, $\tilde{K}$, and $L_2$, may be found in~\cite{BKPS}. 
The analytic expressions of these loop integrals are listed below
\bea
I(\li)&=&\int_{0}^{1}\int_{0}^{1}\frac{dxdy\,y}{B_1(\li)}=-\frac{1}{\lZ}
       \Bigg[ \Li2\left(\frac{1-\li}{1-\li-\lZ\rpl}\right)
       -\Li2\left(\frac{1-\li-\lZ}{1-\li-\lZ\rpl}\right)
       \nonumber\\
& &    +\Li2\left(\frac{1-\li}{1-\li-\lZ\rmi}\right)
       -\Li2\left(\frac{1-\li-\lZ}{1-\lZ-\lZ\rmi}\right)
       \nonumber\\
& &    -\Li2\left(\frac{(1-\li)^2}{(1-\li)^2+\li\lZ}\right)
       +\Li2\left(\frac{(1-\li)(1-\li-\lZ)}{(1-\li)^2+\li\lZ}\right)\Bigg],
       \\
& & \nonumber\\
\tilde{I}(\li)&=&\int_{0}^{1}\int_{0}^{1}\frac{dxdy\,y^2[1-yx(1-x)]}
       {B_1(\li)}=\frac{1}{\lZ}
       \Bigg[\frac{5}{2}-\frac{2(1-\li)}{\lZ}
       \nonumber\\
& &    +2\frac{\li}{\lZ}\ln\li-\frac{2\li}{1-\li}\ln\li
       +4\left(\frac{1-\li}{\lZ}-1\right)\eta\,\tan^{-1}\left( 
       \frac{1}{\eta}\right)
       \nonumber\\       
& &    -\frac{2(1-\li-\lZ)(1-\li)+\li\lZ}{\lZ}I(\li)\Bigg],
       \\
& & \nonumber\\
L_1(\li)&=&\int_{0}^{1}\int_{0}^{1}dxdy\,y
       \ln B_1(\li)=-\frac{3}{2}+\frac{1-\li}{\lZ}
       +\left( 1-\frac{2(1-\li)}{\lZ}\right)\eta\,\tan^{-1}\left(
       \frac{1}{\eta}\right)
       \nonumber\\
& &    -\frac{\li}{\lZ}\ln\li+\frac{(1-\li)^2+\li\lZ}{\lZ}I(\li),
       \\
& & \nonumber\\
K_1(\li,\lj)&=&\int_{0}^{1}\int_{0}^{1}
       \frac{dxdy\,y}{B_2(\li,\lj)}=-\frac{1}{\lZ}
       \Bigg[ \Li2\left(\frac{1-\lj}{1-\lj+\lZ\xpl}\right)
       -\Li2\left(\frac{1-\lj+\lZ}{1-\lj+\lZ\xpl}\right)
       \nonumber\\
& &    +\Li2\left(\frac{1-\lj}{1-\lj+\lZ\xmi}\right)
       -\Li2\left(\frac{1-\lj+\lZ}{1-\lj+\lZ\xmi}\right)
       \nonumber\\
& &    -\Li2\left(\frac{(1-\li)(1-\lj)}{(1-\li)(1-\lj)+\lZ}\right)
       +\Li2\left(\frac{(1-\li)(1-\lj+\lZ)}{(1-\li)(1-\lj)+\lZ}\right)\Bigg],
       \\
& & \nonumber\\
K_2(\li,\lj)&=&\int_{0}^{1}\int_{0}^{1}
       \frac{dxdy\,y^2}{B_2(\li,\lj)}=-\frac{1}{\lZ}
       \Bigg[-1+\frac{1}{1-\li}\ln\li
       \nonumber\\
& &     - \left(\frac{1}{2}-\frac{\li-\lj}{2\lZ}\right)\ln\left(
        \frac{\li}{\lj}\right) 
        +\frac{\sqrt{w}}{\lZ}\tan^{-1}\left(
        \frac{\sqrt{w}}{\li+\lj-\lZ}\right)
       \nonumber\\
& &     +(1-\lj)K_1(\li,\lj)\Bigg]\ +\ (\li\leftrightarrow\lj),
       \\
& & \nonumber\\
\tilde{K}(\li,\lj)&=&\int_{0}^{1}\int_{0}^{1}
       \frac{dxdy\,y^3x(1-x)}{B_2(\li,\lj)}=
       -\frac{1}{\lZ}
       \Bigg[\frac{1}{2}+\frac{2-\li-\lj}{\lZ}
       \nonumber\\
& &    -\frac{1}{\lZ}\ln(\li\lj)
       +\frac{(2-\li-\lj+\lZ)(\lj-\li)}{2\lZ^2}\ln\left(\frac{\li}{\lj}\right)
       \nonumber\\
& &    -\frac{2-\li-\lj}{\lZ}\, \frac{\sqrt{w}}{\lZ}
        \tan^{-1}\left( \frac{\sqrt{w}}{\li+\lj-\lZ}\right)
       \nonumber\\
& &    -\left( 1+\frac{2(1-\li)(1-\lj)}{\lZ}\right) K_1(\li,\lj)\Bigg],
       \\
& & \nonumber\\
L_2(\li,\lj)&=&\int_{0}^{1}\int_{0}^{1}dxdy\,y\ln B_2(\li,\lj)=-\frac{3}{2}
       -\frac{2-\li-\lj}{2\lZ}
       +\frac{1}{2}(\frac{1}{2}+\frac{1}{\lZ})\ln (\li\lj)
       \nonumber\\
& &    +\frac{\li-\lj}{4\lZ^2}(2+2\lZ-\li-\lj)\ln\left( \frac{\li}{\lj}\right)
       +\frac{2-\li-\lj+\lZ}{2\lZ}\nonumber\\
& & \times \frac{\sqrt{w}}{\lZ}
         \tan^{-1}\left( \frac{\sqrt{w}}{\li+\lj-\lZ}\right)
     +\frac{(1-\li)(1-\lj)+\lZ}{\lZ}K_1(\li,\lj),
\eea 
where $\rho_\pm=(1\pm i\eta)/2$ with
      $\eta = \sqrt{4\lZ^{-1}-1}$, 
      $\xi_\pm = (\lZ-\li+\lj\pm i\sqrt{w})/2\lZ$
with  $w=4\li\lj-(\lZ-\li-\lj)^2$, and 
\bea
B_1(\li)&=&(1-y)\li+y[1-\lZ yx(1-x)],
\\
B_2(\li,\lj)&=&1-y+y[x\li+(1-x)\lj-\lZ yx(1-x)].
\eea 
Note that $w\geq 0$ for 
$|\sqrt{\li}-\sqrt{\lj}|\leq\sqrt{\lZ}\leq\sqrt{\li}+\sqrt{\lj}$.
If $\sqrt{\li}+\sqrt{\lj}<\sqrt{\lZ}$,
then one has to analytically continue the function
\bea
\lefteqn{ \sqrt{w}\, \tan^{-1}\left( \frac{\sqrt{w}}{\li+\lj-\lZ}\right)\ =\
2\sqrt{w}\tan^{-1}\left( \sqrt{\frac{\lZ -(\sqrt{\li} -\sqrt{\lj})^2}
{(\sqrt{\li}+\sqrt{\lj})^2-\lZ}} \right)} \nonumber\\
&\to&
\sqrt{-w}\ \ln\left( \frac{\sqrt{1-\frac{(\sqrt{\li}-\sqrt{\lj})^2}{\lZ}}
                     +\sqrt{1-\frac{(\sqrt{\li}+\sqrt{\lj})^2}{\lZ}}}
                     {\sqrt{1-\frac{(\sqrt{\li}-\sqrt{\lj})^2}{\lZ}}
                     -\sqrt{1-\frac{(\sqrt{\li}+\sqrt{\lj})^2}{\lZ}}}\right)
             -i\pi\sqrt{-w}. 
\eea 
The dilogarithmic function $\Li2 (x)$ (with $x$ being real) should also be 
continued analytically as follows:
\beq
\Li2(x\pm i\veps)\ \ =\ \ -\int_0^x dt\frac{\ln |1-t|}{t}\ \pm \ 
i\theta(x-1)\, \pi\ln x .
\eeq 
For $\sqrt{\lZ}<|\sqrt{\li}-\sqrt{\lj}|$, we have checked that Eq.~(A.5) 
and the l.h.s of Eq.~(A.11) do not contain any imaginary part. This implies 
that $\cF^{ll'}_Z$ is pure dispersive in this specific kinematic range. 

As we are interested in heavy neutrinos with masses larger than $M_Z$,
the absorptive part of $\cF_Z^{ll'}$ will solely originate from the Fig.~2(i)
in which only intermediate light neutrinos can come kinematically on-mass shell.
Neglecting light neutrino masses in the calculation, we get
\beq
\Abs(\cF_Z^{ll'})\ =\ i\pi\sum_{i=1}^{\snR}
B_{lN_i}B^\ast_{l'N_i}\; \left[ -\frac{3}{2}
- \frac{1}{\lZ}
+ \left(1+ \frac{1}{\lZ}\right)^2\ln(1+\lZ) \right]. 
\eeq 

\setcounter{equation}{0}
\section{Loop functions of flavour-violating decays \protect\newline
of charged leptons}
\indent

In Section 4 the amplitudes of the flavour-violating decays of $l$,
$l \to l'l_1\bar{l}_2$ and $l\to l'\gamma$, are expressed in terms of 
all possible form factors that are derived by an explicit calculation 
of the Feynman graphs shown in Figs.~1--3.
The photonic form factors, $F_\gamma$ and $G_\gamma$ in 
Eq.~(4.1), vanish in the limit of zero external momenta and lepton
masses due to the electromagnetic gauge invariance. One has consistently
to expand the corresponding loop integrals up to the next order of 
$q^2$~\cite{IL} in order to obtain a nonvanishing result. 
After a straightforward computation, we find that
\bea
F_\gamma(x) &=& \frac{7x^3-x^2-12x}{12(1-x)^3}
               -\frac{x^4-10x^3+12x^2}{6(1-x)^4}\,\ln x,
            \\  
G_\gamma(x) &=& -\frac{2x^3+5x^2-x}{4(1-x)^3}
               -\frac{3x^3}{2(1-x)^4}\,\ln x,
            \\
F_Z(x) &=& -\frac{5x}{2(1-x)}-\frac{5x^2}{2(1-x)^2}\,\ln x,
            \\
G_Z(x,y) &=& -\frac{1}{2(x-y)}\left[\frac{x^2(1-y)}{1-x}\,\ln x
                              -\frac{y^2(1-x)}{1-y}\,\ln y\right],
            \\
H_Z(x,y) &=& \frac{\sqrt{xy}}{4(x-y)}\left[\frac{x^2-4x}{1-x}\,\ln x
                                     -\frac{y^2-4y}{1-y}\,\ln y\right], 
            \\
F_{Box}(x,y) &=& \frac{1}{x-y}
        \Bigg[(1+\frac{xy}{4})\left( \frac{1}{1-x}+\frac{x^2\,\ln x}{(1-x)^2}
                         -\frac{1}{1-y}-\frac{y^2\,\ln y}{(1-y)^2}\right)
            \\
& &         -2xy\left(\frac{1}{1-x}+\frac{x\,\ln x}{(1-x)^2}
             -\frac{1}{1-y}-\frac{y\,\ln y}{(1-y)^2}\right)\Bigg],             
            \nonumber\\
G_{Box}(x,y) &=& -\ \frac{\sqrt{xy}}{x-y}
        \Bigg[(4+xy)\left(\frac{1}{1-x}+\frac{x\,\ln x}{(1-x)^2}
               -\frac{1}{1-y}-\frac{y\,\ln y}{(1-y)^2}\right)
            \nonumber\\
& &      -2\left(\frac{1}{1-x}+\frac{x^2\,\ln x}{(1-x)^2}
           -\frac{1}{1-y}-\frac{y^2\,\ln y}{(1-y)^2}\right)\Bigg].             
\eea 
Although $F_\gamma$, $G_\gamma$, $F_Z$, and $F_{Box}$ are already
known in the literature~\cite{IL,MP,GLR}, the form factors $G_Z$, $H_Z$ and 
$G_{Box}$ are newly obtained by Eqs.~(B.4), (B.5) and (B.7), 
respectively.

For completeness, we list below expressions of the form factors computed 
at some special values of the arguments 
\bea
F_\gamma(1) &=& -\frac{25}{72}, \qquad F_\gamma(0)=0; \qquad
            \\[0.4cm]
G_\gamma(1) &=&  \frac{1}{8},   \qquad G_\gamma(0)=0; \qquad
            \\[0.4cm]
F_Z(1) &=& -\frac{5}{4}, \qquad F_Z(0)=0; \qquad 
            \\[0.4cm]
G_Z(x,x) &=& -\frac{x}{2}-\frac{x\,\ln x}{1-x},\qquad    
G_Z(0,x) = -\frac{x\,\ln x}{2(1-x)},\qquad  G_Z(1,x) = \frac{1}{2},    
         \nonumber \\
G_Z(0,0) &=& 0,  \qquad G_Z(1,0)=\frac{1}{2}, \qquad G_Z(1,1)=\frac{1}{2};
            \\[0.4cm]
H_Z(x,x) &=&  \frac{3}{4}-\frac{x}{4}-\frac{3}{4(1-x)}
             -\frac{x^3-2x^2+4x}{4(1-x)^2}\,\ln x ,  
         \nonumber \\
H_Z(1,x) &=& \frac{\sqrt{x}}{4}\left[ \frac{3}{1-x}-
                           \frac{x^2-4x}{(1-x)^2}\ln x \right],
         \nonumber \\
H_Z(0,x) &=& 0,  \qquad
H_Z(0,0) = 0,  \qquad H_Z(1,0)=0, \qquad H_Z(1,1)=\frac{1}{8};
            \\[0.4cm]
F_{Box}(x,x) &=& -\frac{x^4-16x^3+19x^2-4}{4(1-x)^3}
                 -\frac{3x^3+4x^2-4x}{2(1-x)^3}\,\ln x,
         \nonumber \\
F_{Box}(1,x) &=& -\frac{5x^3-8x^2+7x-4}{8(1-x)^3}  
                 -\frac{x^3-4x^2}{4(1-x)^3}\,\ln x,   
         \nonumber \\
F_{Box}(0,x) &=& \frac{1}{1-x}+\frac{x\,\ln x}{(1-x)^2},  
         \nonumber \\
F_{Box}(0,0) &=& 1, \qquad F_{Box}(1,0)=\frac{1}{2},
                   \qquad F_{Box}(1,1)=\frac{3}{4};
            \\[0.4cm]
G_{Box}(x,x) &=& \frac{2x^4-4x^3+8x^2-6x}{(1-x)^3} 
                -\frac{x^4+x^3+4x}{(1-x)^3}\,\ln x,
         \nonumber \\   
G_{Box}(1,x) &=& -\sqrt{x}\left[\frac{x^3-2x^2+7x-6}{2(1-x)^3}
                          +\frac{x^2-4x}{(1-x)^3}\,\ln x\right],
         \nonumber \\
G_{Box}(0,x) &=& 0, \quad
G_{Box}(0,0) = 0, \quad G_{Box}(1,0)=0, \quad G_{Box}(1,1)=\frac{3}{2}.
\eea 

Since all the form factors given in Eqs.~(B.1)--(B.7) are multiplied 
by certain combinations of $B$ and $C$ matrices in the decay 
amplitudes~(4.1), (4.6), (4.7) and (4.8), it will be helpful to 
define the following composite form factors:
\bea
F_\gamma^{l l'} &=& \sum_{i} B^\ast_{li}B_{l'i}F_\gamma(\li)
                 =  \sum_{N_i} B^\ast_{lN_i}B_{l'N_i}F_\gamma(\lNi),
            \\
G_\gamma^{l l'} &=& \sum_{i} B^\ast_{li}B_{l'i}G_\gamma(\li)
                 =  \sum_{N_i} B^\ast_{lN_i}B_{l'N_i}G_\gamma(\lNi),
            \\
F_Z^{l l'} &=& \sum_{ij} B^\ast_{li}B_{l'j}
     \Big[\delta_{ij}F_Z(\li)+C^\ast_{ij}G_Z(\li,\lj)+C_{ij}H_Z(\li,\lj)\Big]
            \nonumber\\
           &=&\sum_{N_iN_j} B^\ast_{lN_i}B_{l'N_j}
           \Big[\delta_{N_iN_j}(F_Z(\lNi)+2G_Z(0,\lNi))
            \nonumber\\     
           & &+C^\ast_{N_iN_j}(G_Z(\lNi,\lNj)-G_Z(0,\lNi)-G_Z(0,\lNj))
            \nonumber\\     
           & &+C_{N_iN_j}H_Z(\lNi,\lNj)\Big],
            \\
F_{Box}^{l l' l_1 l_2} &=& 
  \sum_{ij} B^\ast_{li} B^\ast_{l_2j}(B_{l'i}B_{l_1j}+B_{l_1i}B_{l'j})
    \: F_{Box}(\li,\lj)
            \nonumber\\
  & & +\ \sum_{ij} B^\ast_{li} B^\ast_{l_2i}B_{l'j}B_{l_1j}\: G_{Box}(\li,\lj) 
            \nonumber\\
  &=& \sum_{N_iN_j}\Bigg[( B^\ast_{lN_i}B_{l'N_i}\delta_{l_1 l_2}
                    +B^\ast_{lN_i}B_{l_1N_i}\delta_{l' l_2})
                     \delta_{N_iN_j}\Big[ F_{Box}(0,\lNi)-F_{Box}(0,0)\Big]
            \nonumber\\
  & & +B^\ast_{lN_i} B^\ast_{l_2N_j}(B_{l'N_i}B_{l_1N_j}+B_{l_1N_i}B_{l'N_j})
            \nonumber\\
  & &              \times\Big[ F_{Box}(\lNi,\lNj)-F_{Box}(0,\lNj)
                     -F_{Box}(0,\lNi)+F_{Box}(0,0)\Big]
            \nonumber\\
  & & +B^\ast_{lN_i} B^\ast_{l_2N_i}B_{l'N_j}B_{l_1N_j} G_{Box}(\lNi,\lNj)
               \Bigg],
\eea 
where we have made use of the identities of Eq.~(2.9)
in the final step of the Eqs.~(B.15)--(B.18) and re-expressed all
the composite form factors as a sum over the heavy neutrino states.
This simplification enables us to study the behaviour of these
form factors in the heavy neutrino limit. 

For the purpose of illustration, we will discuss the results of this 
asymptotic limit in a model with two heavy Majorana neutrinos. Employing 
Eqs.~(5.1) and~(5.2) for the mixing matrices $B$ and $C$, we find that for
$\lN1= m^2_{N_1}/M^2_W \gg 1$ and $\rho=m^2_{N_2}/m^2_{N_1} \gg 1$,
\bea
F_\gamma^{l l'} &\to & -\ \frac{1}{6}\; s_L^{\nu_l}s_L^{\nu_{l'}} \ln\lN1, 
   \\
G_\gamma^{l l'} &\to & \frac{1}{2}\; s_L^{\nu_l}s_L^{\nu_{l'}},
\eea 
\bea
F_Z^{l l'} &\to & -\; \frac{3}{2}s_L^{\nu_l}s_L^{\nu_{l'}}\ln\lN1\; 
               \nonumber\\
           && +\; s_L^{\nu_l}s_L^{\nu_{l'}}\sum_{i=1}^{\snG}\ (s_L^{\nu_i})^2\;
                  \frac{\lN1 }{(1+\rho^{\frac{1}{2}})^2}\left(-\frac{3}{2}\rho
                   +\frac{-\rho+4\rho^{\frac{3}{2}}-\rho^2}
                         {4(1-\rho)}\ln\rho\right)\;
               ,\\
F_{Box}^{l l'l_1l_2}&\to &-\; (s_L^{\nu_l}s_L^{\nu_{l'}}\delta_{l_1l_2}
                           +s_L^{\nu_l}s_L^{\nu_{l_1}}\delta_{l'l_2}) 
               \nonumber\\
            && +\; s_L^{\nu_l}s_L^{\nu_{l'}}s_L^{\nu_{l_1}}s_L^{\nu_{l_2}}\;
               \frac{\lN1 }{(1+\rho^{\frac{1}{2}})^2}
               (-\rho-\frac{\rho+\rho^{\frac{3}{2}}+\rho^2}
                           {1-\rho}\ln\rho).
\eea 
In the limit $\rho \to 1$ and for $\lN1\gg 1$, Eqs.~(B.21) and~(B.22) take 
the form
\bea
F_Z^{l l'}& \to &-\; \frac{3}{2}s_L^{\nu_l}s_L^{\nu_{l'}}\ln\lN1\; 
                -\; \frac{1}{2} s_L^{\nu_l}s_L^{\nu_{l'}}\sum_{i=1}^{\snG}\ 
                             (s_L^{\nu_i})^2\lN1
               ,\\
F_{Box}^{l l'l_1l_2}&\to &
               -\; (s_L^{\nu_l}s_L^{\nu_{l'}}\delta_{l_1l_2}
                +s_L^{\nu_l}s_L^{\nu_{l_1}}\delta_{l'l_2})\; 
               +\; \frac{1}{2}s_L^{\nu_l}s_L^{\nu_{l'}}s_L^{\nu_{l_1}}
                s_L^{\nu_{l_2}}\; \lN1.
\eea 
From Eqs.~(B.19)--(B.24), it is obvious that all the composite form factors, 
$F_\gamma^{l l'}$, $G_\gamma^{l l'}$, $F_Z^{l l'}$, and $F_{Box}^{l l'l_1l_2}$, 
violate the decoupling theorem~\cite{AC}. Note that terms of 
$\cO ((s^{\nu_l}_L)^2)$ in $F^{ll'}_Z$ depend
logarithmically on the heavy neutrino mass, $m_{N_1}$, while terms 
proportional to $(s^{\nu_l}_L)^4$ in Eqs.~(B.23) and~(B.24) show a strong 
quadratic, $m^2_{N_1}/M^2_W$, dependence and should not be neglected in the 
calculation when $m_{N_1}> 200$~GeV (see, {\em e.g.}, Fig.~4). 

\setcounter{equation}{0}
\section{Three-body phase-space integrals}
\indent

As we have seen from Eq.~(4.6), the $\gamma$-mediated amplitude of
the decay $l(p)\to l'(p')l_1(p_1)\bar{l}_2(p_2)$ contains a non-local
interaction which leads to a collinear singularity in the limit
$q^2\equiv (p_1+p_2)^2 \to 0$. This divergency can only be avoided
if one assumes that the leptons, $l_1$ and $l_2$, coupled to the
virtual photon are not strictly massless, {\it i.e.} 
$m_{l_1}=m_{l_2}=\veps\ne 0$. Thus, after
performing phase-space integration, we neglect all those terms 
that vanish as $\veps\to 0$. On the other hand, the mass of $l'$ can safely 
be set to zero.
Then, the phase-space boundaries can be given by
\beq
4\veps^2\ \leq\ s_2\ \leq\ m^2, \qquad
s_1^\pm\ =\ \frac{1}{2}(m^2-s_2)\left[\, 1\pm \sqrt{1-\frac{4\veps^2}
{s_2}}\, \right]\, +\, \veps^2,
\eeq 
where $s_1=(p'+p_2)^2$, $s_2=(p_1+p_2)^2$, $m$ is the mass of the decaying
lepton $l$, and $s^{+(-)}_1$ is the upper (lower) limit of the Mandelstam 
variable $s_1$. 

The divergent phase-space integrals relevant for the decay $l\to l'l_1
\bar{l}_2$ (with $l'\neq l_2$) are given by the following expressions: 
\bea
P_1 &=&\int\int ds_2ds_1\frac{1}{s_2}
     = m^2(\ln\frac{m^2}{\veps^2}-3)+\cO(\veps),
     \\
P_2 &=&\int\int ds_2ds_1\frac{s_1}{s_2}
     = m^4(\frac{1}{2}\ln\frac{m^2}{\veps^2}-\frac{7}{4})+\cO(\veps),
     \\
P_3 &=&\int\int ds_2ds_1\frac{s_1^2}{s_2}
     = m^6(\frac{1}{3}\ln\frac{m^2}{\veps^2}-\frac{4}{3})+\cO(\veps).
\eea 
Note that a different result would have been obtained in Eqs.~(C.2)--(C.4) 
if we had originally expanded the square root existing in $s_1^\pm$ 
in terms of $\veps$ and then performed the phase-space integration. 
Apparently, this technical problem seems to have 
caused some confusion in the literature, as far as the 
correct analytic expression of the non-local interaction in Eqs.~(4.9) 
and~(4.10) is concerned. In the three-body leptonic decays of $l$
where $l'=l_1=l_2$, one may have to take into account an additional
divergent phase space integral when interfering the two possible,
$s_1$-channel and $s_2$-channel, $\gamma$-exchange amplitudes, {\it i.e.}
\beq
P_4 \ =\ \int\int ds_2ds_1\frac{1}{s_2s_1}
     =-\frac{\pi^2}{12}-\ln^22+\frac{1}{2}\ln^2\frac{m^2}{\veps^2}
     +\cO(\veps).
\eeq 
The integral $P_4$ in Eq.~(C.5), however, is always multiplied by the
small mass $\veps$ of the final leptons and therefore goes to zero
as $\veps\to 0$. As a result, the only type of divergency for 
$\veps\to 0$ that appears in Eqs.~(4.9) and~(4.10) is the logarithmic one, 
$\ln(m^2/\veps^2)$. 

\end{appendix}

\newpage


\newpage

\centerline{\Large{\bf Figure Captions} }
\vspace{-0.2cm}
\newcounter{fig}
\begin{list}{\rm{\bf Fig. \arabic{fig}:} }{\usecounter{fig}
\labelwidth1.6cm \leftmargin2.5cm \labelsep0.4cm \itemsep0ex plus0.2ex }

\item Feynman graphs responsible for generating the effective vertex 
$\gamma l l'$ ($l\neq l'$).

\item Feynman graphs responsible for generating the effective vertex 
$Z l l'$ ($l\neq l'$).

\item Feynman diagrams relevant for the leptonic decays 
$l\to l'l_1 \bar{l}_2$.

\item $B(\tau^- \to e^-e^-e^+)$ (solid line) and 
$B(\tau^- \to e^-\mu^-\mu^+)$ (dashed line)
as a function of the heavy neutrino mass $m_N(=m_{N_1}=m_{N_2})$ assuming
$(s^{\nu_\tau}_L)^2=0.07$, $(s^{\nu_e}_L)^2=0.015$ and 
$(s^{\nu_\mu}_L)^2\simeq 0$. Numerical results obtained
when seemingly suppressed terms of $\cO ((s^{\nu_l}_L)^4)$ are neglected,
are also presented for $B(\tau^- \to e^-e^-e^+)$ (dotted line) and 
$B(\tau^- \to e^-\mu^-\mu^+)$ (dash-dotted line).

\item $B(\tau^- \to e^-e^-e^+)$ (solid line) 
and $B(\tau^- \to e^-\mu^-\mu^+)$ (dashed line)
as a function of the heavy neutrino mass $m_N(=m_{N_1}=m_{N_2})$ using
$(s^{\nu_\tau}_L)^2=0.035$, $(s^{\nu_e}_L)^2=0.010$ and 
$(s^{\nu_\mu}_L)^2\simeq 0$. We also display numerical results 
obtained by neglecting seemingly suppressed terms of $\cO ((s^{\nu_l}_L)^4)$ 
in the calculation of $B(\tau^- \to e^-e^-e^+)$ (dotted line) and 
$B(\tau^- \to e^-\mu^-\mu^+)$ (dash-dotted line).

\item $B(\tau^- \to e^-e^-e^+)$ as a function of the ratio $m_{N_2}/m_{N_1}$ 
for $m_{N_1}=200$~GeV (solid line) and 500~GeV (dashed line).  
We have assumed $(s^{\nu_\tau}_L)^2=0.07$, $(s^{\nu_e}_L)^2=0.015$ and 
$(s^{\nu_\mu}_L)^2\simeq 0$. The corresponding numerical results
of $B(\tau^- \to e^-\mu^-\mu^+)$ are shown for $m_{N_1}= 200$~GeV
(dotted line) and 500~GeV (dash-dotted line).

\item Numerical estimates of $B(Z \to e^-\tau^+)+B(Z \to e^+\tau^-)$ 
as a function of the heavy neutrino mass $m_N(=m_{N_1}=m_{N_2})$ for 
three representative values of the mixing parameters 
$(s^{\nu_\tau}_L)^2$ and $(s^{\nu_e}_L)^2$ ($(s^{\nu_\mu}_L)^2=0$): 
(i) $(s^{\nu_\tau}_L)^2=0.070$ and $(s^{\nu_e}_L)^2=0.015$ (solid line),
(ii) $(s^{\nu_\tau}_L)^2=0.035$ and $(s^{\nu_e}_L)^2=0.010$ (dashed line),
and (iii) $(s^{\nu_\tau}_L)^2=0.020$ and $(s^{\nu_e}_L)^2=0.010$ (dotted line).

\item Numerical estimates of $B(Z \to e^-\tau^+)+B(Z \to e^+\tau^-)$ 
versus $m_{N_2}/m_{N_1}$ for selected values of 
$m_{N_1}=200$~GeV (solid line),
400~GeV (dashed line), 600~GeV (dotted line), and 1~TeV (dash-dotted 
line). \nopagebreak
We have used $(s^{\nu_\tau}_L)^2=0.07$ and $(s^{\nu_e}_L)^2=0.015$.

\end{list}

\end{document}